# Fundamental Curie Temperature Limit in Ferromagnetic $Ga_{1-x}Mn_xAs$

K. M. Yu and W. Walukiewicz,
Electronic Materials Program, Materials Sciences Division,
Lawrence Berkeley National Laboratory, Berkeley, CA 94720

T. Wojtowicz,
Institute of Physics, Polish Academy of Sciences, Warsaw, Poland and Department of
Physics, University of Notre Dame, Notre Dame, IN 46556

W.L. Lim, X. Liu, U. Bindley, M. Dobrowolska, and J. K. Furdyna
Department of Physics, University of Notre Dame, Notre Dame, IN 46556

ABSTRACT

We provide experimental evidence that the upper limit of $\sim$110 K commonly observed for the Curie temperature $T_C$ of $Ga_{1-x}Mn_xAs$ is caused by the Fermi-level-induced hole saturation. Ion channeling, electrical and magnetization measurements on a series of $Ga_{1-x-y}Mn_xBe_yAs$ layers show a dramatic increase of the concentration of Mn interstitials accompanied by a reduction of $T_C$ with increasing Be concentration, while the free hole concentration remains relatively constant at $\sim5x10^{20}cm^{-3}$. These results indicate that the concentrations of free holes and ferromagnetically active Mn spins are governed by the position of the Fermi level, which controls the formation energy of compensating interstitial Mn donors.

The recent discovery of III-V ferromagnetic semiconductors, specifically $Ga_{1-x}Mn_xAs$ with Curie temperatures $T_C$ as high as 110 K [1-2] is a major step toward the implementation of spintronic devices for processing, transferring, and storing of information [3]. Experimentally it has been established that $T_C$ in $Ga_{1-x}Mn_xAs$ increases with increasing Mn concentration x (as long as MnAs precipitates are not formed) and with hole concentration. Although each Mn atom in the Ga sublattice is expected to contribute a hole to the system, it was found that the hole concentration in this material is significantly lower than the Mn concentration (by a factor of 2-3) [1,2]. Recent ion channeling experiments demonstrated that such low Mn acceptor activation could be attributed to the presence of *interstitial Mn donors* in $Ga_{1-x}Mn_xAs$ [4].

Calculations based on the Zener model [5] predicted that $T_C$ in $Ga_{1-x}Mn_xAs$ could be improved by increasing the Mn content and/or the free hole concentration in the alloy. These predictions led to extensive experimental works aimed at achieving higher $T_C$ for $Ga_{1-x}Mn_xAs$. Despite intense efforts, similar maximum values of $T_C$ of ~110K were found in thin $Ga_{1-x}Mn_xAs$ films prepared in different laboratories with rather different values of *x*, ranging from ~0.05 to 0.10 and optimally annealed at low temperatures in the range of 250-280ºC [2, 6-9]. A recent report by Potashnik *et al.* showed that in optimally annealed $Ga_{1-x}Mn_xAs$ alloys, the $T_C$ and conductivity saturate for x> 0.05 [8], suggesting that as x increases, an increasing fraction of Mn spins do not participate in the ferromagnetism.

In an earlier post-growth annealing study of $Ga_{0.91}Mn_{0.09}As$ films [4,9] we found that annealing at 280ºC for one hour increases $T_C$ from 65 K to 111 K and the hole concentration from $6x10^{20}$ $cm^{-3}$ to $1x10^{21}$ $cm^{-3}$. Ion channeling results demonstrated that this increase of both $T_C$ and the hole concentration can be attributed to the lattice site rearrangement of the highly unstable Mn interstitials $Mn_I$. These $Mn_I$ are expected to be highly mobile positively charged double donors [10,11]. They can, however, be immobilized by occupying the interstitial sites *adjacent* to the negatively-charged substitutional Mn acceptors ($Mn_{Ga}$), thus forming antiferromagnetically ordered $Mn_I$-$Mn_{Ga}$ pairs, which not only render $Mn_{Ga}$ inactive as acceptors, but also cancels its magnetic moment [4,10]. Low temperature annealing breaks up the relatively weak antiferromagnetically ordered $Mn_I$-$Mn_{Ga}$ pairs, leading to a higher concentration of uncompensated Mn spins, resulting in increase in saturation magnetization as well as a higher hole concentration and a higher $T_C$ [4,6-9].

The above low temperature annealing results further suggested the possibility that there exists a *fundamental limit* on $T_C$, governed by a limit on the hole concentration allowed by the $Ga_{1-x}Mn_xAs$ alloy [4]. In this paper we use co-doping of $Ga_{1-x}Mn_xAs$ by Be as a tool to provide unambiguous experimental evidence that such a limit does in fact exist. It has been demonstrated that free hole concentration as high as $8 \times 10^{20}$ cm$^{-3}$ could be achieved in Be doped, low temperature grown GaAs [12]. We show that the free hole concentration $p$ in $Ga_{1-x-y}Mn_xBe_yAs$ with x=0.05 is *nearly constant*, independent of the Be doping level (up to y = 0.11). In spite of this *saturation* of $p$, we observe for a fixed Mn concentration of 0.05 a dramatic increase in the concentrations of $Mn_I$ and of electrically inactive random Mn clusters at the expense of $Mn_{Ga}$ as the Be concentration is increased, accompanied by *a strong decrease of $T_C$.* These results strongly indicate that a Fermi-level-controlled mechanism puts an upper limit on $T_C$ in $Ga_{1-x}Mn_xAs$ [13].

Thin films of $Ga_{1-x-y}Mn_xBe_yAs$ were grown on semi-insulating (001) GaAs substrates in a Riber 32 R&D MBE system. Prior to film deposition we grew a 450 nm GaAs buffer layer at 590ºC (i.e., under normal GaAs growth conditions). The substrate was then cooled down for the growth of a 3 nm thick low-temperature (LT) GaAs, followed by a 230 nm thick layer of $Ga_{1-x-y}Mn_xBe_yAs$ at a substrate temperature of 270ºC. The $As_2$:Ga beam equivalent pressure ratio of 20:1 was maintained during the growth.

Magnetoresistance, Hall effect, and SQUID magnetometry were used for electrical and magnetic characterization of the samples and for determining $T_C$. Hall effect measurements were performed in the Van der Pauw or the six-probe geometry. To circumvent the problems associated with the *anomalous Hall effect* (AHE) in ferromagnets [2, 14], we have used the electrochemical capacitance voltage (ECV) profiling method to measure the depth distribution of acceptors in our specimens. By comparing the Hall and ECV results on non-ferromagnetic $Ga_{1-y}Be_yAs$ thin films grown under similar conditions as the $LT\text{-}Ga_{1-x}Mn_xAs$ and $Ga_{1-x-y}Be_yMn_xAs$ films, we have established that ECV can be reliably used to obtain the free hole concentration profiles in ferromagnetic $LT\text{-}Ga_{1-x}Mn_xAs$ [15].

The locations of Mn sites in the $Ga_{1-x}Mn_xAs$ lattice were studied by simultaneous channeling particle induced x-ray emission (c-PIXE) and Rutherford backscattering spectrometry (c-RBS) using a 1.95MeV $^4He^+$ beam. Mn $K_\alpha$ x-ray signals obtained by c-PIXE are directly compared with GaAs c-RBS signals coming from $Ga_{1-x}Mn_xAs$ films. The normalized yield for the RBS ($\chi_{GaAs}$) or the PIXE Mn x-ray signals ($\chi_{Mn}$) is defined as the ratio of the channeled yield to the corresponding unaligned "random" yield.

Figure 1 shows the PIXE and RBS angular scans (normalized yield as a function of the tilt angle around the channeling axis) about the <110> (taken along the {110} planar direction) and <111> axes for the $Ga_{1-x}Mn_xAs$ and $Ga_{1-x-y}Be_yMn_xAs$ films with increasing y (results from only four out of six samples were shown for simplicity). The angular scans about the <100> directions are similar to those about the <111> direction for all samples and are therefore not shown. The total Mn content in all samples was determined by PIXE to be ~0.05. The Be contents was estimated from the lattice constant determined by x-ray diffraction that was calibrated by RHEED intensity oscillations.

For all the samples studied, the <111> axial Mn scans (c-PIXE) follow the host GaAs (RBS) scans, indicating that the dominant fraction of the Mn atoms are either on substitutional sites or are on specific sites shadowed by the host atoms [16,17]. This reveals that the majority of the Mn atoms is on specific (non-random) sites commensurate with the lattice, but does not necessarily imply that all of the Mn atoms are in *substantial* positions. At the same time the normalized yields $\chi_{Mn}$ in the <111> scans also shows a gradual increase, deviating from the corresponding host scans as the Be content increases, indicating an increase in Mn atoms in the form of random clusters not commensurate with the GaAs lattice.

In contrast to the <111> angular scans, the Mn <110> angular scans are strikingly different from their corresponding host scans in Fig. 1. In the sample without Be (y = 0), we observe that the <110> $\chi_{Mn}$ is significantly higher than that in the <111> scan, particularly in the middle of the channel, suggesting that a significant fraction of the non-random Mn shadowed in the <111> scans do not all occupy substitutional sites, and can thus be assumed to be located at the *interstitial* sites lying along the <111> axis of the zinc-blende crystal lattice. Atoms in these interstitial positions, tetrahedral or hexagonal in a diamond cubic lattice are shadowed by the host atoms when viewed along both the <100> and <111> axial directions. They are, however, exposed in the <110> axial channel [16,17], giving rise to a double-peak (tetrahedral site) or a single peak (hexagonal site) feature in the <110> angular scan due to the flux-peaking effect of the ion beam in that channel [16]. We find from the difference between the <110> and <111> scans for this sample that the fraction of Mn in interstitial sites amounts to ~7%.

As the Be content increases, the <110> Mn angular scans show a definite peak at the center of the channel that increases in intensity -- a clear signature for the presence of an increasing concentration of Mn interstitials in the alloy [18]. These results unambiguously reveal that the fraction of $Mn_I$ as well as random Mn-related clusters increases monotonically in $Ga_{1-x-y}Be_yMn_xAs$ films with increasing Be content. The fractions of Mn atoms at the various sites -- substitutional ($Mn_{Ga}$), interstitial ($Mn_I$) and in random-cluster form ($Mn_{ran}$) -- as measured from the angular scans are shown in Fig. 2. Mn atoms in various lattice locations for a samples with y~0.03 and 0.08 and annealed at 280ºC for 1 hr. are also shown. Notice that when the samples are annealed a dramatic increase of Mn as random clusters at the expense of $Mn_I$ is observed while the $Mn_{Ga}$ fraction stays the same revealing the relative instability of the $Mn_I$.

Figure 3 shows the free hole concentration obtained from ECV and Hall measurements together with the Curie temperature $T_C$ for as-grown samples with different Be content y. It is particularly worth noting that $T_C$ of the $Ga_{1-x-y}Mn_xBe_yAs$ films drops rapidly as y increases – in fact the samples become non-ferromagnetic for y > 0.05 -- while the free hole concentration measured by ECV remains rather constant throughout the entire Be composition range. We point out that for the ferromagnetic $Ga_{1-x}Mn_xAs$ thin film where the $T_C$ is high a large discrepancy in the hole concentration measured by Hall effect is observed due to the strong AHE even at room temperature. As the Be concentration in the film increases (y > 0.05), the $Ga_{1-x-y}Mn_xBe_yAs$ films lose their ferromagnetic property and the hole concentrations measured by Hall effect is seen to approach that measured by the ECV method.

The ECV data show that the different $Ga_{1-x-y}Be_yMn_xAs$ films have similar values of free hole concentration of ~4 to $6x10^{20}$/cm$^3$. It has been established that in compound semiconductors the carrier concentration is limited by the formation of compensating native defects. The formation energies of these defects are governed by the position of the Fermi level [19,20]. The relatively constant hole concentration of about $5x10^{20}$/cm$^3$ shown in Fig. 3 indicates that the hole concentration in these $Ga_{1-x-y}Mn_xBe_yAs$ samples is at the free hole saturation limit $p_{max}$. As this limit is reached, the formation energies of $Mn_{Ga}$ acceptors and compensating $Mn_I$ become comparable. Introduction of additional Be acceptors into the $Ga_{1-x-y}Mn_xBe_yAs$ samples then leads to a downward shift of the Fermi energy, that in turn increases the formation energy of negatively charged $Mn_{Ga}$ acceptors. As a result, an increasing fraction of Mn is incorporated in the form of $Mn_I$ donors and/or electrically inactive MnAs or Mn clusters [21]. The creation of $Mn_I$ not only puts a limit on the maximum hole concentration, but also has a profound effect on the number of ferromagnetically active spins and -- for a constant hole concentration -- on the RKKY coupling of these spins.

Specifically, there are three mechanisms to note in this context. First, it has been shown theoretically that $Mn_I$ on tetrahedral sites do not participate in the RKKY-mediated ferromagnetism because $Mn_I$ $d$-orbitals do not hybridize with the $p$-states of the holes at the top of the valence band [10]. Second, as mentioned earlier, the $Mn_I$ donors may form antiferromagnetically ordered $Mn_I$-$Mn_{Ga}$ pairs [10], which not only renders $Mn_{Ga}$ inactive as acceptors, but also reduces the total number of *uncompensated Mn spins participating in the ferromagnetism.* Such drop in the number of active spins reduces $T_C$. Finally, when the number of active spins becomes approximately equal to the hole concentration, the average distance between the active Mn spins becomes larger than the first node in the oscillatory RKKY exchange coupling (at $\approx 1.17$ $r_{hole}$, where $r_{hole}$ is the average distance between holes [22]). In this situation some $Mn_{Ga}$ ions may couple antiferromagnetically between themselves. This would at first lead to the drop in $T_C$, and eventually should drive the system into a spin-glass state [22-24]. We believe that some or all of the above factors contribute to the strong drop in $T_C$ and to the disappearance of ferromagnetism in $Ga_{1-x-y}Mn_xBe_yAs$ with increasing Be content.

In conclusion, our present work on LT- $Ga_{1-x-y}Mn_xBe_yAs$ alloys, together with previously reported studies of the low temperature annealing of $Ga_{1-x}Mn_xAs$, reveal that the ferromagnetism in $Ga_{1-x}Mn_xAs$ is related to the total number of uncompensated Mn ions, which are in turn controlled by the formation energies of compensating native defects. As the Mn concentration x increases beyond the doping limit $p_{max}$, it is energetically favorable to form compensating $Mn_I$, thus keeping the product of the free hole concentration and of the concentration of the net uncompensated Mn spins participating in the ferromagnetism relatively constant at the maximum level. Given that the ferromagnetism in this system is related to the uncompensated Mn spins and is mediated by holes, such Fermi-level-induced hole saturation effect necessarily imposes a fundamental limit on the Curie temperature of the system. Since the total number of acceptors has to be maintained below $p_{max}$, co-doping of $Ga_{1-x}Mn_xAs$ with Be acceptors creates a huge increase of $Mn_I$ , thus killing ferromagnetism. This experimental observation leads us to propose using *heavy n-type counter-doping* of $Ga_{1-x}Mn_xAs$ (with, e.g., Te) as a remedy for the otherwise unavoidable creation of Mn interstitials at higher values of x. In such $Ga_{1-x}Mn_xTe_zAs_{1-z}$ it should be possible to achieve values of $x \approx p_{max}+z$. Although the hole concentration will still be "pinned" at $p_{max}$ by the limit imposed on the Fermi level, the number of active Mn would increase in proportion to x, thus increasing $T_C$.

This work was supported by the Director, Office of Science, Office of Basic Energy Sciences, Division of Materials Sciences and Engineering, of the U. S. Department of Energy under Contract No. DE-AC03-76SF00098; by NSF Grant DMR00-72897; and by the DARPA SpinS Program.

as 230 hours. Surface and/or interface effects are expected to play an important role in this case. Some improvements on $T_C$ have been also achieved by Ohno et. al in Be co-doped double layer structures. However, considerations of the thermodynamic equilibrium in such nonuniform systems are beyond the scope of this article.

FIGURE CAPTION

Fig. 1   Angular scans about the <110> and <111> axes for undoped and for Be-doped $Ga_{1-x}Mn_xAs$ samples.  The <110> angular scans are taken along the {110} planar channel.

Fig. 2   The fractions of Mn atoms at the various sites -- substitutional ($Mn_{Ga}$), interstitial ($Mn_I$) and in random-cluster form ($Mn_{ran}$)-- as measured from the angular scans shown in Fig. 1.  The Mn fractions for the sample with y~0.08 annealed at 280ºC for 1 hr. are also shown as closed symbols.

Fig. 3   The hole concentrations determined by ECV and Hall measurements and the Curie temperatures $T_C$ for as-grown $Ga_{1-x-y}Mn_xBe_yAs$ films with increasing Be content y.

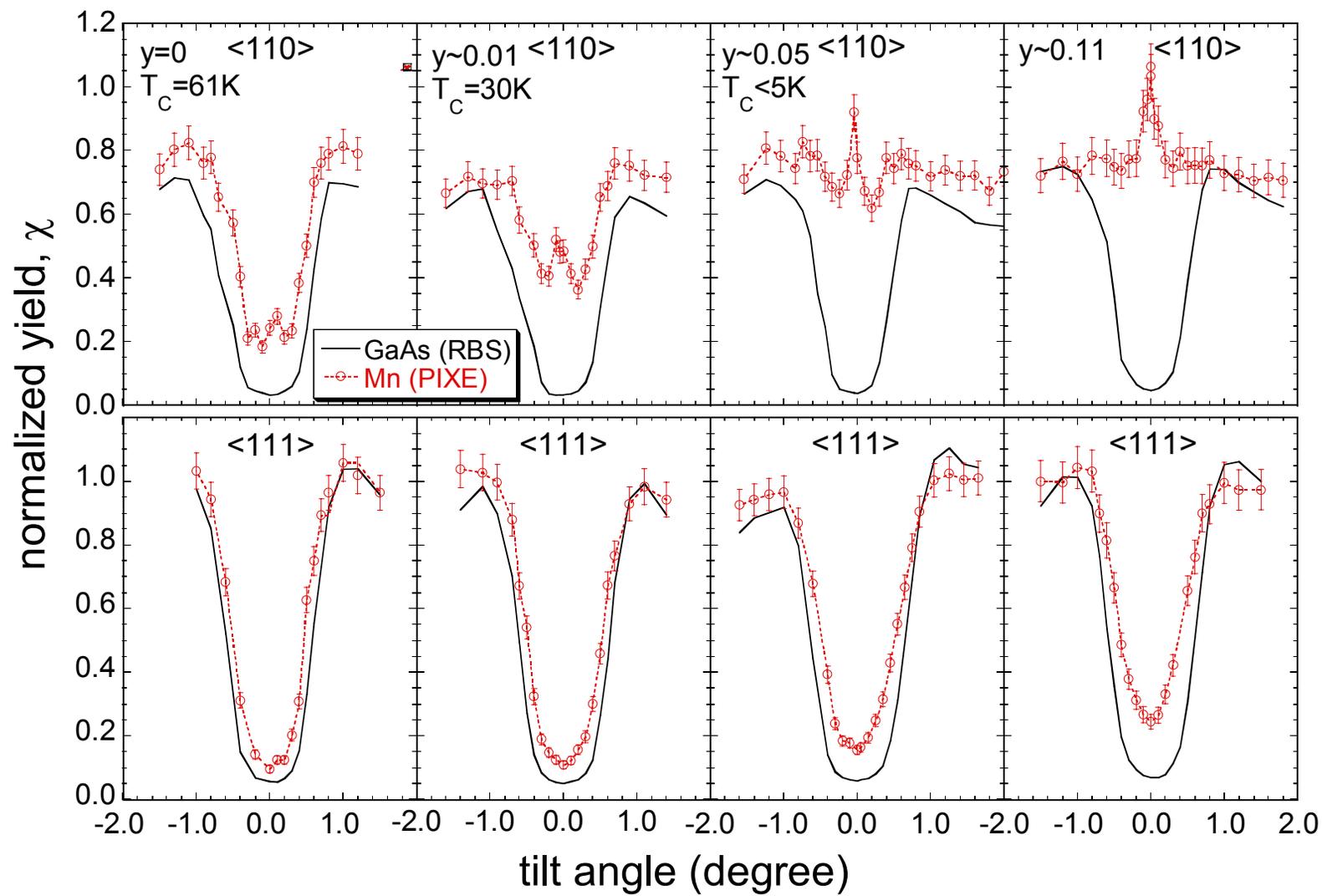

Fig. 1

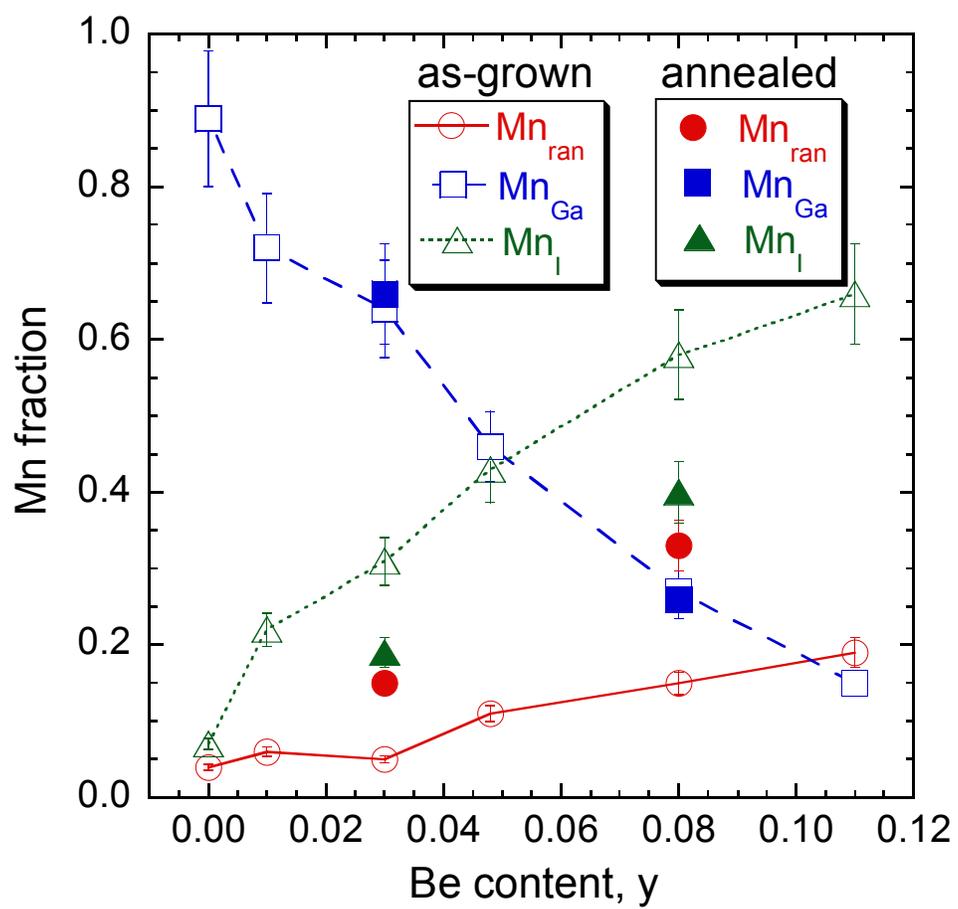

Fig. 2

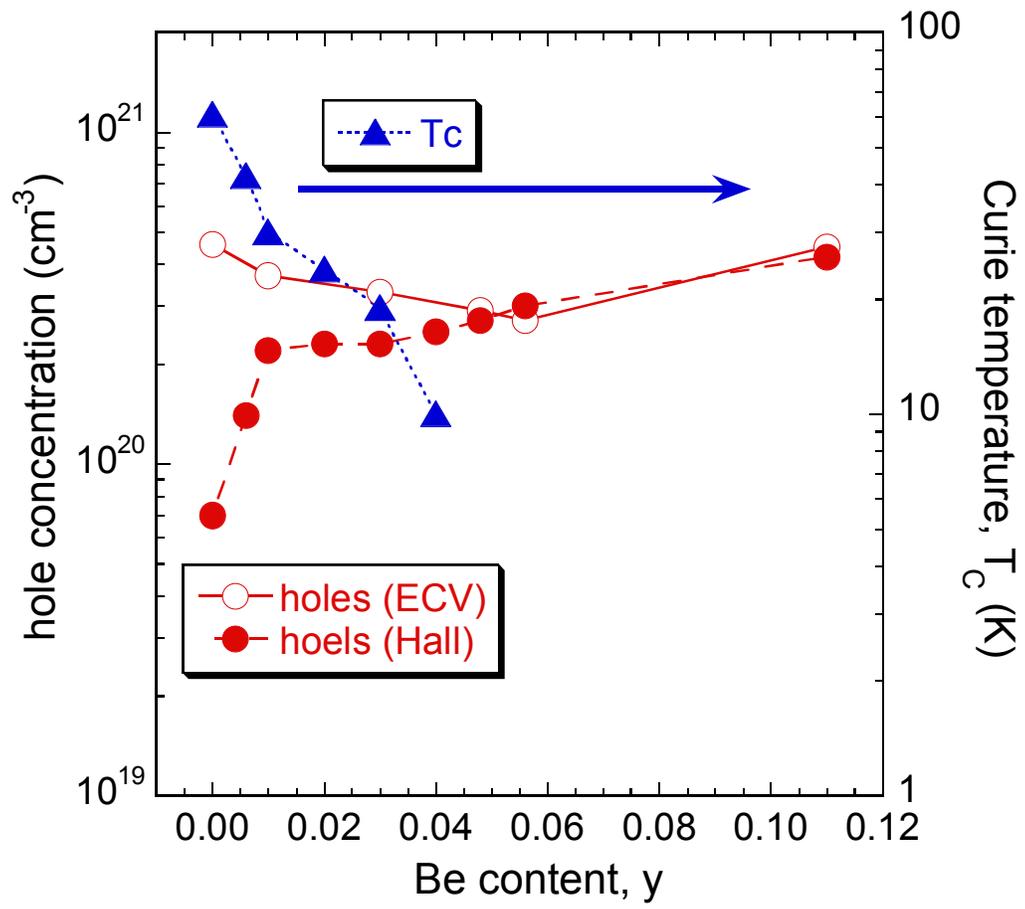

Fig. 3